\documentclass[traditabstract]{aa}

\usepackage{epsfig,amsmath,amssymb}

\def\msol{M_\odot}
\def\sig{ \sigma }
\def\ga{\,\hbox{\hbox{$ > $}\kern -0.8em \lower 1.0ex\hbox{$\sim$}}\,}
\def\la{\,\hbox{\hbox{$ < $}\kern -0.8em \lower 1.0ex\hbox{$\sim$}}\,}
\def\beq{\begin{equation}}
\def\eeq{\end{equation}}

\titlerunning{Dimensional argument for turbulent support}
\authorrunning{Chabrier and Hennebelle}

\begin{document}

\title{Dimensional argument for the impact of turbulent support on the stellar initial mass function}

\author{Gilles Chabrier\inst{1,2} \and Patrick Hennebelle\inst{3}}

\institute{CRAL (UMR CNRS 5574), Ecole Normale Sup\'erieure de Lyon, 69364 Lyon Cedex, France 
\and
School of Physics, University of Exeter, Exeter, EX4 4QL, UK 
\and
LERMA (UMR CNRS 8112), Ecole Normale Sup\'erieure, 75231 Paris Cedex, France}

\abstract{
We present a simple dimensional argument to illustrate the impact of nonthermal support from turbulent velocity dispersion on the shape of the 
 prestellar core mass function (CMF), precursor of the stellar initial mass function (IMF).
The argument demonstrates the need to invoke such support  to recover the Salpeter slope in the high-mass part of the CMF/IMF, whereas
pure thermal support leads to a much steeper slope. This simple dimensional argument clearly highlights the results obtained in the complete Hennebelle-Chabrier theory of the IMF.
}


\maketitle

\section{Introduction}

Understanding the shape of the stellar initial mass function (IMF) has been an ongoing domain of research in astrophysics since the pioneering work of Salpeter (1955). 
Indeed, the IMF is the fundamental parameter governing stellar and galaxy evolution, and it is rooted in the very process of star formation. 
In the modern context of star formation, it is now widely accepted that star formation is a dynamical process, where turbulence plays a major role 
(see e.g Brunt 2003, MacLow \& Klessen 2004, McKee \& Ostriker 2007). 
Turbulence is injected on large scales in the ISM by various processes of stellar or galactic origin so that turbulent motions determine the structure of all density regimes 
in the interstellar gas  and in star-forming molecular clouds (Elmegreen \& Scalo 2004). 
In weak density media, like molecular clouds, such large-scale supersonic turbulence cascades to small scales by shocks.

Turbulence is by nature a nonlinear process, developing from the strong distortions of the fluid velocity field due to advection
(mathematically formalized by the $({\bf u}\cdot {\bf \nabla})$ operator in the hydrodynamic equations). Interactions between these strong random distortions of the velocity field globally act as a nonthermal source of
support, stabilizing otherwise unstable scales. Often, and incorrectly for large-scale turbulence, such a support is denominated for convenience nonthermal "pressure".

The statistics of the gas density fluctuations created by turbulence is captured by the so-called probability density function (PDF). It is generally
admitted that, in good approximation, the PDF of turbulence resembles a lognormal form, as expected from the application of the central limit theorem to
 a hierarchical density field generated by multiplicative processes, such as
shocks (V\'azquez-Semadeni 1994). Such a lognormal behaviour for density fluctuations has received observational support. Indeed, observationally, the 3D lognormal shape of the PDF is reflected by the projected (2D)
power spectrum column density of molecular clouds, measured from dust extinction maps (Kainulainen et al. 2009, Lombardi et al. 2010, Brunt et al. 2010).  
Determining the {\it velocity} field fluctuations is more problematic.  As one can normally only measure
the velocity dispersion along the line of sight (${\sigma}_{z}$), it is only
possible to constrain the tangential velocity dispersion
by assuming isotropy.

In the case of compressible turbulence, the standard deviation of the logarithm of the density fluctuations $\log (\rho/{\bar \rho}) $, i.e. the width of the PDF, determined from numerical simulations, is found to obey reasonably well the relation (e.g. Padoan \& Nordlund 2002, Federrath et al. 2010)

\beq
\sigma_{\log(\rho)}^2 \simeq \ln [1+(b{\cal M})^2],
\label{sig}
\eeq
where $\cal M$ is the hydrodynamical rms Mach number. This corresponds to a standard deviation $\sigma_{\rho} \simeq  b{\cal M}$ for the density fluctuation
$\rho/{\bar \rho}$. The factor $b$ is found to be $b\approx 0.5$ in 3D simulations for a mixture
of solenoidal and compressive turbulence driving modes (Federrath et al. 2010, Schmidt et al. 2010), in agreement with observational determinations (Padoan, Jones \& Nordlund 1997, Brunt 2010). Interestingly enough, the density PDF and relation (\ref{sig}) seem to remain weakly affected,
for ${\cal M} \la 10$,
in the case of supersonic MHD turbulence (Lemaster \& Stone 2008, Kritsuk et al. 2011, Price et al. 2011).

The first comprehensive theory of the IMF in the context of turbulent fragmentation of a molecular cloud has been formalized by Padoan \& Nordlund (2002, PN). In this theory small-scale cascade of large-scale turbulence by shocks generates overdense regions that may become
gravitationally unstable, giving birth to the prestellar cores. A different theory has been proposed more recently by Hennebelle \& Chabrier (2008, 2009, HC; see
Hennebelle \& Chabrier 2011 for a brief summary of these two theories). Although both theories predict that turbulence promotes small-scale overdensities,
triggering the formation of low-mass objects, in particular brown dwarfs, the two theories differ on several points. A fundamental difference, in particular, concerns the role of
nonthermal support on the determination of the IMF.
 On one hand, PN do not include turbulent support in their theory, and must invoke magnetic field and MHD shock conditions to obtain the proper Salpeter slope for the IMF. In contrast, in the HC theory, random turbulent motions inside bound massive overdense regions, identified at the very early stages of star formation, counteract the action of gravity and thus stabilize these initial mass reservoirs - precursors of the gravitationally bound prestellar "cores" - which would otherwise have collapsed before collecting such an amount of mass. As such, these turbulent motions are
acting like a source of nonthermal support (identified in the HC theory by the quantity ${\cal M}_\star$, the Mach number at the Jeans scale, see Eqs.(41) and (43) of HC08 ) against gravity.
As demonstrated in HC08, this turbulent support is mandatory to obtain the proper Salpeter slope.
However, by the time the (massive) cores/stars form out of the (massive) reservoirs,
turbulence will have had ample time to dissipate. We come back to this important concept of turbulent "support" at the end of the paper.

Conversely, a fundamental prediction of the HC theory is that, in the absence of non-thermal support, the slope of the IMF is significantly steeper than the Salpeter one.
This result, hence the importance of the aforementioned source of nonthermal support in order to get the proper Salpeter slope for the IMF, has received some support
from recent simulations aimed at exploring this issue (Schmidt et al. 2010), even though more work is definitely needed to nail down this issue.

In this short paper, we present a simple dimensional argument, based on the well-known scaling properties of turbulence, to demonstrate that indeed, turbulent
support is needed to recover the proper Salpeter high-mass tail of the CMF/IMF.

\section{Scaling properties of turbulence}
\label{scale}

The average statistical properties of a turbulent density and velocity field can be described by the scale dependence of the corresponding 3D
power spectra in Fourier space, as described by power-law relationships, $P_{3D}(x)\equiv P_x(k)\propto k^{-n_x}$,  where $x$ denotes either the velocity ($x \equiv\sigma$),
density ($x \equiv \rho/{\bar \rho})$) or {\it logarithm} of density ($x \equiv \log(\rho/{\bar \rho}) $) fluctuations.
The energy spectrum, E(k), which is the kinetic energy per unit mass per unit wavenumber, 
is equal to the angular average of the power spectrum: $E(k) =4\pi k^2 \langle P_{3D}(k)\rangle  \propto k^{(-n+2)}$ in 3D.
It corresponds to the spectrum of velocity fluctuations over scales
smaller than a size $ R $, $\sigma_R^2  =\int_{2\pi/R}^\infty E(k)dk \propto \int_{2\pi/R}^\infty  k^{(-n+2)}dk \propto R ^{n-3}$, where
$\sigma_ R ^2$ describes the variance of the 3D velocity field, $\sigma(x, y, z)$, within a volume $R ^3$. In real space,
the energy spectrum thus corresponds to the second-order rms velocity fluctuations  measured on scale $R$,  $\langle  \sigma_R ^2 \rangle$.

The two well known limits of  incompressible and pressureless turbulence correspond to the Kolmogorov and Burgers limits, with respective 3D velocity power spectrum indexes $n=11/3$ and $n =4$.
Numerical simulations of compressible, shock-dominated turbulence under Mach values typical of molecular cloud conditions tend to show that the power spectrum of turbulent  velocity and {\it logarithm} of density fluctuations exhibit a power-law behaviour, with a similar index, namely $n_\sigma \sim n_{\log(\rho)} \sim 3.8$ (Kritsuk et al. 2007).

\section{Analytical dimensional argument}
\label{dim}

According to the Hennebelle-Chabrier theory, in the random field of density fluctuations characteristic of the cloud's structure, initial pieces of fluid out of which prestellar cores and individual stars (or multiple star systems such as binaries) will eventually form, are identified as overdense regions that isolate themselves from the surrounding
medium and start to contract under the action of gravity. The condition to select such a given marginally unstable piece of fluid of mass $M$ and size $R$,  with internal velocity dispersion $\sigma_R\equiv\sig$, simply stems from the virial condition: $2E_{kin}+E_{pot}=0$, 
where $E_{kin}\sim (1/2)\rho \sig^2 $ denotes
the kinetic energy and $E_{pot}\sim \rho G\frac{M}{R}$ the gravitational energy. Using dimensional analysis, this yields the relation

\begin{eqnarray}
 M\sig^2\sim G \frac{M^2}{R}{\hskip .5cm}{\rm i.e.} {\hskip .5cm}M\sim {R\sig^2 \over G}\,.
\label{vir}
\end{eqnarray}

\noindent In the limit where the velocity dispersion is dominated by thermal motions, and assuming nearly isothermal conditions, $\sig$ does not depend on $R$ and condition (\ref{vir}) yields
\begin{eqnarray}
 M_{th}\propto R{\hskip 2.8cm}({\rm THERMAL\,\,SUPPORT}).
\label{therm}
\end{eqnarray}

\noindent Conversely, in the limit where the velocity dispersion is dominated by turbulent motions,  $\sig \propto R^{{(n-3)\over 2}}$ (see \S\ref{scale}), yielding for the marginal instability condition

\begin{eqnarray}
M_{turb} \propto  R^{(n-2)} \propto R^{2\eta+1}{\hskip .2cm}({\rm TURBULENT\,\,SUPPORT}).
 \label{Mturb}
\end{eqnarray}
with $\eta=(n-3)/2$. 
The thermal case is recovered for $n=3$, i.e. $\eta=0$. For $n>3$, i.e. $\eta>0$, we note the stronger dependence of the mass $M$ upon the size $R$ in the presence of turbulent support, compared with the
pure thermal case (Eq.(\ref{therm})). The complete relation is obtained in HC08 (their eqn.(26)) and illustrates that a bound stable region can achieve larger masses for a given size
thanks to the non-thermal support at the characteristic Jeans length, naturally leading to a larger number of high mass reservoirs (then of massive prestellar cores), which would otherwise have collapsed long before. This impact of nonthermal support on the mass-size relation of massive bound cores has been confirmed by recent
numerical simulations (Schmidt et al. 2010, Fig. 7).

\bigskip

 Assuming a large enough multi-scale structured space of size $L$ and  dimension $D$, thus volume $L^D$, 
 the probability of having $N$ gravitationally bound structures on a scale larger than $R$ 
is given by the probability law ${\cal P}(N_{R})\equiv N(>R)\propto R^{-D}$.
In Fourier space, this simply means that the number of fluctuations of wavenumber $k\sim 1/R$ is proportional to the volume of a fluctuation, i.e. $dN(k)\propto d{\vec k}=k^{D-1}dk$.
The density probability, i.e. the probability to have a structure of size $\in[R,R+dR]$, thus reads as
\begin{eqnarray}
\frac{dN(R)}{dR}\propto R^{-(1+D)}. \label{N}
\end{eqnarray}
A homogeneous volume in 3 dimensions obviously corresponds to $D=3$, i.e. a uniform density probability $dN(k)/d{\vec k}=constant$, as assumed e.g. in Padoan \& Nordlund (2002).

From Eqs.~(\ref{Mturb}) and (\ref{N}), the probability of having a bound structure of mass $\in[M(R),M+dM(R+dR)]$, which defines the CMF/IMF, reads
\begin{eqnarray}
\frac{dN}{dM}&\propto& M^{-\alpha}\nonumber \\
{\rm with}\,\,\,\alpha&=&{n+1 +(D-3) \over n-2} ={2\eta+4 +(D-3) \over 2\eta+1} 
\label{Nturb}
\end{eqnarray}

\noindent 
\noindent As seen from Eq.(\ref{Nturb}), for $D=3$, pure thermal support ($n=3,\eta =0$) yields a power-law IMF $\frac{dN}{dM}\propto M^{-4}$, substantially steeper than the 
Salpeter value, $\alpha=2.35$\footnote{Assuming a Gaussian (lognormal) density probability instead of a uniform one yields in that case $\frac{dN}{dM}\propto M^{-3}$ (Eq.(37) of HC08).}.
In the two limiting cases of incompressible (Kolmogorov) ($n=11/3,\eta =1/3$) and pressureless (Burgers) ($n=4,\eta =1/2$) turbulence, the turbulent support yields
$\alpha=2.8$ and $\alpha=2.5$, respectively. In molecular clouds, turbulence is assumed to scale according to the observed Larson (1981) velocity dispersion - size relation
with a typical value $\eta\sim 0.4$-0.5. This value is well recovered for the aforementioned power spectrum index inferred from numerical simulations of supersonic turbulence, $n=3.8$.
According to eqn.(\ref{Nturb}), this yields a slope for the CMF/IMF $\alpha=2.66$,
between the Kolmogorov and Burgers values.

There are observational suggestions, however, that molecular clouds have a fractal dimension (Elmegreen \& Falgarone 1996), with $D\approx 2.5$-2.7 (S\'anchez et al. 2005). 
Herschel observations of gravitationally bound prestellar cores, on the other hand, suggest  a mass-size relation $M\propto R^\beta$, with $\beta\approx 1$-2 (Fig. 4 of Andr\'e et al. 2010 and  K{\" {o}}nyves et al. 2010). 
Combined with a Salpeter mass distribution, this yields $dN/dR \propto R^{-(1+D)}$ with $D\approx 1.4$-2.7. Although these results should be taken with caution, they suggest
a structured space with $D<3$ all the way from clouds to cores. This is consistent with turbulence being the dominant mechanism that {\it initially} determines the distribution of both unbound and bound structures. It arises, however, from two different properties of turbulence.
As demonstrated in HC08, the structure of the {\it unbound} clouds/clumps arises from the (log){\it density} structure of turbulence, characterized by a lognormal PDF and a power spectrum index $n^\prime$. On the other hand, the distribution of 
gravitationally {\it bound} structures is set up by the virial condition and thus involves the properties of the {\it velocity} structure of turbulence (see Eq.~(\ref{vir})),
characterized by a power spectrum index $n$. As mentioned above, both indexes appear to be similar for supersonic turbulence, with $n^\prime\simeq n\simeq 3.8$ (Kritsuk et al. 2007).
Although the mass-size relation of (evolved) prestellar cores cannot be directly applied to the one characteristic of the initial mass reservoirs, it is interesting, however, to note that a value $n=3.8$ in Eq.~(\ref{Mturb}) yields a relation for the bound structures supported by turbulence $M\propto R^{1.8}$ (see Eq.(43) and below in HC08 for the
complete relation),  consistent with the
aforementioned observational determination from Herschel.

Interestingly, using the aforementioned observationally inferred value $D\approx 2.5$-2.7 yields $\alpha=3.5$-3.7, according to Eq.(\ref{Nturb}),  in the case of purely thermal support, which is
still significantly steeper than the Salpeter value, $\alpha=2.25$-2.35 and 2.5-2.6 in the Burgers and Kolmogorov limits, respectively, 
and $\alpha=2.39$-2.5 when using the aforementioned value for the supersonic turbulence index.

\section{Discussion and conclusion}
\label{conclusion}

The complete calculations, as done in HC08 and HC09, yield of course more accurate determinations of the slope of the CMF, both in the thermal and nonthermal case.
It is important to stress that, in the present simplistic approach, one {\it assumes} a space distribution given by Eq.~(\ref{N}),
 which leads to the ($D$-$3$) correction in Eq.~(\ref{Nturb}).
In the rigorous (Press-Schechter based)
statistical approach developed in HC08 and HC09, the dimension $D$, both for the (unbound) clumps and the (bound)
cores is {\it naturally encapsulated in the (density and velocity) scaling properties of turbulence} entering the HC theory.
There is no need to explicitly assume a
fractal index, so there is no explicit dependence of the Salpeter slope upon such an index.
On the contrary, the size distribution naturally arises from the calculations with (as easily inferred from eqn.(42) of HC08 and the mass-size relation (\ref{Mturb}) - detailed in HC08 -
derived from the viral condition):
$ dN/dR \propto R^{-{n+3\over 2}}\approx R^{-3.4}$, 
i.e. $D\approx 2.4$, consistent with the aforementioned Herschel determinations.
In contrast, a pure thermal support (eqn.(\ref{therm}) and $n=3$) 
in the complete HC theory yields $ dN/dR \propto  R^{-{3n-3\over 2n-4}}\propto R^{-3}$, i.e. $D=2$.
The sometimes advocated dependence of the IMF upon the spatial dimension $D$, and the fractional value of the latter, thus
simply reflect the (stabilizing) impact of the turbulent velocity field on large-scale bound structures. 

The simple
dimensional relations derived in this short note thus
enable us to grasp the crucial impact of turbulent support, combined with gravity through the virial condition, on the shape
of the IMF. Indeed, based on simple scaling arguments, these relations demonstrate
that nonthermal support is mandatory to yield the correct high-mass power-law part of the IMF. Pure thermal support would produce a much steeper IMF, leading to a
drastic lack of massive ($\ga 1\,\msol$) cores/stars. Interestingly, Eq.(\ref{Nturb}) yields $d\alpha/dn<0$, as in HC, and thus predicts that the IMF should {\it steepen}
with a {\it decreasing} velocity power spectrum index. This could be tested
with dedicated numerical simulations.  It also shows that magnetic field is not needed to recover the power-law slope of the IMF. 

As mentioned in Hennebelle \& Chabrier (2008), the limitation of the HC theory is the lack of time dependence. Indeed, initial massive mass
reservoirs that have been selected, in our statistical formalism, as being stabilized by turbulence
might in reality fragment into smaller pieces during the cloud's evolution.
The impact of such a time dependence on the IMF is presently under study. Further fragmentation of these primordial cores, however, is unlikely to significantly affect the high-mass part of the IMF, for two reasons. First of all, given the scale-invariance of turbulence, although  fragmentation of massive cores during evolution may affect the lower mass part of the IMF, by producing new small cores, it is unlikely to drastically affect the high-mass part, as turbulence-induced fragmentation processes will be similar for all masses. 
Statistical explorations of core fragmentation indeed show that the high-mass part of the IMF is still barely affected (Swift \& Williams 2008).
Second of all, observations of massive cores suggest that fragmentation is rather limited, most of the mass of the core ending up in one or just a few smaller cores (Bontemps et al. 2010; Longmore et al. 2011). Indeed, various numerical simulations of collapsing dense cores show that radiative feedback and magnetic fields drastically reduce the fragmentation process (e.g. Krumholz et al. 2007, Commer\c con et al. 2010, Hennebelle et al. 2011).

\bigskip
Together with the results obtained in recent dedicated numerical
simulations (Schmidt et al. 2010), these simple dimensional arguments strongly argue in favor of the high-mass part  of the initial core mass function, hence of the resulting IMF, being mainly determined by the presence of hydrodynamical turbulent motions at the early stages of star formation.
It is essential, however, to clearly understand what such a "turbulent support" corresponds to in the present approach (i.e. in the Hennebelle-Chabrier theory). As briefly outlined in the Introduction, turbulent support here is not to be understood in a static sense, as often (most of the time incorrectly) in the literature, by supposing that turbulence will remain present in the core and will act permanently as an extra source of pressure. Indeed, turbulence within an overdense region will decay within about a crossing time. The mass-scale relation
stated by eq.(\ref{Mturb}) will thus eventually evolve into a thermal relation, as given by eq.(\ref{therm}), as turbulence dissipates during the collapse of the core. As
mentioned earlier, however, the aforederived relations apply to the {\it initial mass reservoirs} set up by the density fluctuations within the cloud/clump. Turbulent support in that sense means that the random motions of the flow provide enough kinetic energy {\it at the early stages of structure formation} for massive reservoirs to circumvent the action of gravity, thus
stabilizing these otherwise collapsing structures. In other words, overdense regions with lower masses than the {\it turbulent} Jeans mass (see Eqs.(25-26) of HC08 and eq.(\ref{Mturb}) above) will be dispersed back into the flow under the action of turbulence within a turbulent crossing time, before they become dominated by gravity. Only in regions exceeding the turbulent Jeans mass will gravity take over, yielding the collapse of the corresponding mass reservoirs, providing the initial seeds for the massive gravitationally bound prestellar cores. In that sense, the concept of a turbulent Jeans mass should be seen as a statistical criterion for selecting the initial pieces of gas that will be dominated by gravity, and will thus collapse, from the ones that will be dispersed, within a turbulent crossing time.

\begin{acknowledgements}
The authors are grateful to the referee, A. Whitworth, whose remarks greatly helped improve the manuscript.
The research leading to these results received funding from the European Research Council under the European Community's Seventh Framework Programme (FP7/2007-2013 Grant Agreement no. 247060).

\end{acknowledgements}

\section*{Appendix}
A point is worth mentioning concerning the slope of the CMF/IMF. Let us reason out in 3D, for sake of simplicity. In 3D, the case $D=3$ corresponds in the k-space to a structured space such that $N(k)\propto k^3$, i.e. to a space filling distribution of fluctuations (maximum compactness). Since the mass associated with the fluctuation of
scale $R$ is $M=\rho R^3\propto k^{-3}$, one immediately gets
\begin{eqnarray}
\frac{dN}{dM}\propto M^{-2}.
\label{geo}
\end{eqnarray}
The mass spectrum (\ref{geo}) thus corresponds to the most natural mass distribution, set up by fluctuations obeying purely geometrical considerations, i.e. a white noise spectrum of fluctuations. It is thus not surprising that mass spectra in the interstellar medium exhibit a power-law behaviour with an exponent close to 2, namely $\sim 1.6$ for unbound structures such as CO clumps, $\sim 2.0 $ for stellar clusters, and $\sim 2.35$ for cores/stars. Therefore, the aim of any theory of the IMF is to explain the departure from the naturally expected $\alpha=2$ slope, which arises from physical, not purely geometrical, effects.


\begin{thebibliography}{}
\bibitem[]{} Andr\'e et al. 2010, \aap, 518, L102
\bibitem[]{} Bontemps, S., Motte, F., Csengeri, T., \& Scheider, N., 2010, \aap, 524, 18
\bibitem[Brunt 2003]{B03} Brunt, C., 2003, \apj, 584, 293
\bibitem[Brunt et al. 2010]{Bruntetal10} Brunt, C., Federrath, C., Price, D., 2010, \mnras, 405, L56
\bibitem[Brunt 2010]{B10} Brunt, C. 2010, \aap, 513, 67
\bibitem[]{} Commer\c con, B., Hennebelle, P., Audit, E., Chabrier, G., Teyssier, R., 2010, \aap, 510, L3
\bibitem[Elmegreen \& Falgarone]{} Elmegreen \& Falgarone, 1996, \apj, 471, 816
\bibitem[Elmegreen \& Scalo 2004]{ES04} Elmegreen, B., Scalo, J., 2004, \araa, 42, 211
\bibitem[Federrath et al. 2010]{F10} Federrath, C., Roman-Duval, J., Klessen, R. S., Schmidt, W., Mac Low, M.-M., 2010, \aap, 512, 81
\bibitem[Hennebelle \& Chabrier 2008]{HC08} Hennebelle, P., Chabrier, G., 2008, \apj, 684, 395 (HC08)
\bibitem[Hennebelle \& Chabrier 2009]{HC09} Hennebelle, P., Chabrier, G., 2009, \apj, 702, 1428 (HC09)
\bibitem[Hennebelle \& Chabrier 2011]{HC11} Hennebelle, P., Chabrier, G., 2011, in {\it Computational Star Formation}, Proceedings of the International Astronomical Union, IAU Symposium, 270, 159
\bibitem[]{} Hennebelle, P., Commer\c con, B., Joos, M., Klessen, R., Krumholz, K., Tan, J., \& Teyssier, R., 2010, \aap, 528, 72
\bibitem[Kainulainen et al. 2009]{K09} Kainulainen, J., Lada, C., Rathborne, J., Alves, J., 2009, \aap, 497, 399
\bibitem[]{} K{\" {o}}nyves et al., 2010, \aap, 518, L106
\bibitem[Kritsuk et al. 2007]{Kritsuketal2007} Kritsuk, A., Norman, M., Padoan, P., Wagner, R., 2007, \apj, 665, 416
\bibitem[Kritsuk et al. 2011]{Kritsuketal2011} Kritsuk, A., Ustyugov, S., Norman.,  2011, in {\it Computational Star Formation}, Proceedings of the International Astronomical Union, IAU Symposium, 270, 179
\bibitem[]{} Krumholz, M., Klein, R., McKee, C., 2007, \apj, 656, 959
\bibitem[Larson 1981]{Larson81} Larson, R., 1981, MNRAS, 194, 809
\bibitem[Lemaster \& Stone 2008]{LS08} Lemaster, M., Stone, J., 2008, \apj, 682, L97
\bibitem[]{} Longmore, S. N., Pillai, T., Keto, E., Zhang, Q.;,Qiu, K., 2010, \apj, 726, 97
\bibitem[Lombardi et al. 2010]{L10} Lombardi, M., Lada, C. J., Alves, J., 2010, \aap, 512, A67
\bibitem[MacLow \& Klessen 2004]{MacLowKlessen04} MacLow, M.-M.,  Klessen, R., 2004, Rev. Mod. Phys., 76,
\bibitem[McKee \& Ostriker 2007]{McKO07} McKee, C., Ostriker, E., 2007, \araa, 45, 565
\bibitem[Padoan et al. 1997]{Padoanetal97} Padoan, P., Jones, B., Nordlund, A., 1997, \apj, 474, 730
\bibitem[Padoan \& Nordlund 2002]{PadoanNordlund02} Padoan, P., Nordlund, A., 2002, \apj, 576, 870 (PN)
\bibitem[Price et al. 2011]{Price11} Price, D., Federrath, C., Brunt, C., 2011, \apj, 727, L21
\bibitem[Salpeter 1955]{S55} Salpeter, E., 1955, \apj, 121, 161
\bibitem[S\'anchez05]{} S\'anchez, N, Alfaro, E., P\'erez, E., 2005, 625, 849
\bibitem[Schmidt et al. 2010]{F10} Schmidt, W., Kern, S., Federrath, C., Klessen, R., 2010, \aap, 516, 25
\bibitem[Swift \& Williams 2008]{SW08} Swift, J., \& Williams, J., 2008, \apj, 679, 552
\bibitem[V\'azquez-Semadeni 1994]{VS94} V\'azquez-Semadeni, E., 1994, \apj, 423, 681
\end{thebibliography}
\end{document}